\documentclass{sig-alternate-05-2015}

\usepackage{tabularx}

\begin{document}

% Copyright
\setcopyright{acmcopyright}

% DOI
\doi{10.475/123_4}

% ISBN
\isbn{123-4567-24-567/08/06}

%Conference
\conferenceinfo{SWDM2016}{October 28, 2016, Indianapolis, USA}

\acmPrice{\$15.00}

\title{LeaveNow: A Social Network-based Smart Evacuation System for Disaster Management}

\numberofauthors{4}
\author{
% 1st. author
\alignauthor
Ziyuan Wang\\
       \affaddr{IBM Research - Australia}\\
       \affaddr{Melbourne, Australia}\\
       \email{ziwang@au1.ibm.com}
% 2nd. author
\alignauthor
Jianbin Tang\\
       \affaddr{IBM Research - Australia}\\
       \affaddr{Melbourne, Australia}\\
       \email{jianbi.tang@au1.ibm.com}
% 3rd. author
\alignauthor 
Yini Wang\\
       \affaddr{Neusoft}\\
       \affaddr{Beijing, China}\\
       \email{wangyini@neusoft.com}
\and  % use '\and' if you need 'another row' of author names
% 4th. author
\alignauthor 
Bo Han\\
       \affaddr{Hugo Australia Pty Ltd}\\
       \affaddr{Sydney, Australia}\\
       \email{bhan@hugo.ai}
\alignauthor 
Xi Liang\\
       \affaddr{Deakin University}\\
       \affaddr{Melbourne, Australia}\\
       \email{xi.liang@deakin.com.au}
}

\maketitle

\begin{abstract}
The importance of timely response to natural disasters and evacuating affected people to safe areas is paramount to save lives. Emergency services are often handicapped by the amount of rescue resources at their disposal. We present a system that leverages the power of a social network forming new connections among people based on \textit{real-time location} and expands the rescue resources pool by adding private sector cars. We also introduce a car-sharing algorithm to identify safe routes in an emergency with the aim of minimizing evacuation time, maximizing pick-up of people without cars, and avoiding traffic congestion.
\end{abstract}


%\ccsdesc[500]{Computing methodologies~Massively parallel and high-performance simulations}


%
%  Use this command to print the description
%
\printccsdesc


\section{Introduction}
Natural disasters such as fire and floods cause global loss of life every year. Fast and effective evacuations are crucial to reduce the death toll in these emergency situations. However, rescue teams are often restricted by the limited resources (e.g., the number of evacuation vehicles, helicopters) on hand, not allowing them to respond timely to all the affected people. To bridge the gap, we propose a social network-enabled LeaveNow system that solicits private car resources from social sectors, e.g., volunteers to complement official rescue resources. It aggregates requests for help and available cars based on real-time locations and coordinates car-sharing services during the evacuation.  Unlike existing social networks, LeaveNow \textit{connects people in real-time based on location}. Thanks to the car-sharing service and routing optimization empowered by backend services, LeaveNow supports coordinative evacuation actions to mitigate traffic congestion and enable an efficient use of available resources during evacuation. To our knowledge, existing systems~\cite{pang2013selecting,pourrahmani2015dynamic} do not offer such features.

We further propose an algorithm which ensures safety of volunteer drivers and aims to save more lives. The algorithm identifies all exits in the danger zone based on a road map. Each volunteer driver is automatically advised to take the exit that minimizes their own evacuation time. The algorithm also maximizes the number of passengers assigned to volunteer drivers to be picked up on their evacuation routes. Both volunteers and passengers benefit from such scheduling as the road traffic is largely reduced. 

\section{LeaveNow System}
\begin{figure}[!htbp]
\centering
\includegraphics[width=7.5cm]{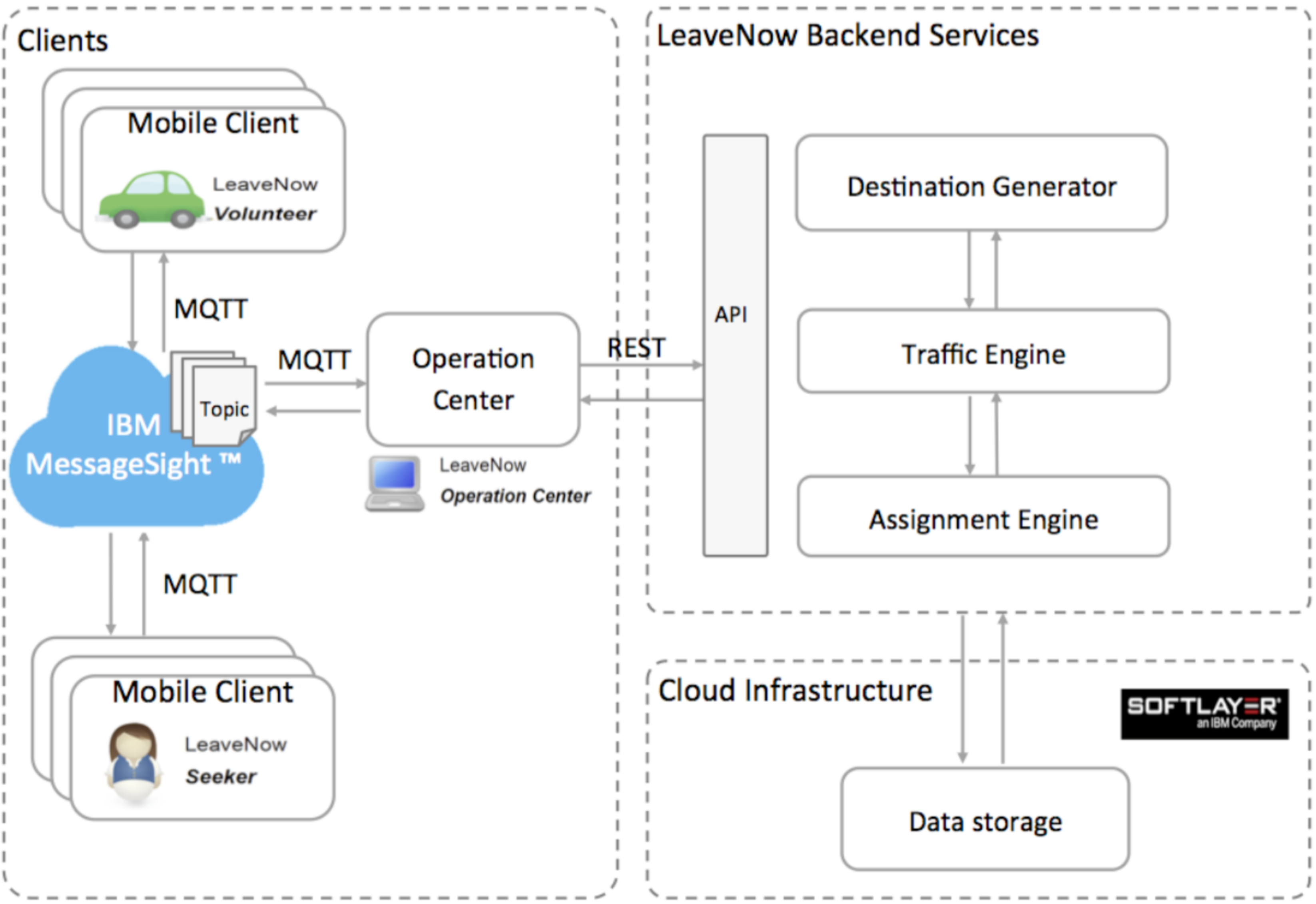}
\caption{LeaveNow system architecture}
\label{fig:arch}
\vspace{-8pt}
\end{figure}

In this section, we present the LeaveNow system architecture, including the rescue-and-routing algorithm, and describe the system workflow and user interfaces. 
The LeaveNow system consists of three layers as shown in Figure \ref{fig:arch}: 

\textbf{Clients:} \textit{Seeker} is a mobile client interface, used by people who require car-sharing services if they lack the means to leave the disaster affected area in a timely manner.
\textit{Volunteer} is another mobile client interface used by volunteer drivers who offer pick-up services for the seekers.
\textit{Operation Centre} provides several functionalities, including: a) broadcasting warnings of danger; b) receiving the location and number of service seekers and volunteer cars; c) aggregating and transmitting information to the back-end services; and d) monitoring the progress of the current evacuation.

\textbf{Backend Services:} This layer is the central coordinator. \textit{Destination Generator} generates all exit locations of the specified danger zone (geographic boundary). Danger zones can be manually selected by a qualified operator or automatically determined by interpreting the affected area using information shared in the social network or broadcasted via official channels. Destination Generator applies the Multi-Agent Transport Simulation Toolkit  (MATSim) \footnote{http://www.matsim.org/} to automatically identify the location of all exits, which is the intersection of the danger zone and all segments of road map in MATSim. 
\textit{Traffic Engine} generates route plans on the basis of road maps, traffic conditions and evacuation help requests. Traffic Engine computes the evacuation routes of volunteers based on the travel
distance to each exit, speed limits and traffic congestion. For each volunteer driver, Traffic Engine generates the best safe route to each exit from its current location. Among these routes, the exit that has the shortest travel time is assigned to the volunteer.
\textit{Assignment Engine} evaluates and selects an optimized evacuation scheduling determined by a pickup and routing algorithm. Each volunteer has a corresponding exit determined by Traffic Engine. Seekers are assigned to volunteers, provided they are within a pre-defined distance to the route. Volunteers then pick up seekers on their way to exits.

\textbf{Cloud Infrastructure:} It offers an Infrastructure-as-a-Service (IaaS), which enables clients to access the  \textit{Data Storage} and other computing resources. We leverage SoftLayer Object Storage to store the input (road segment map, evacuation zone, traffic simulator configuration) and output (exits, evacuation routes and seeker assignment schema) of the whole LeaveNow system.

\begin{figure}[!hp]
\centering
\includegraphics[width=6.5cm]{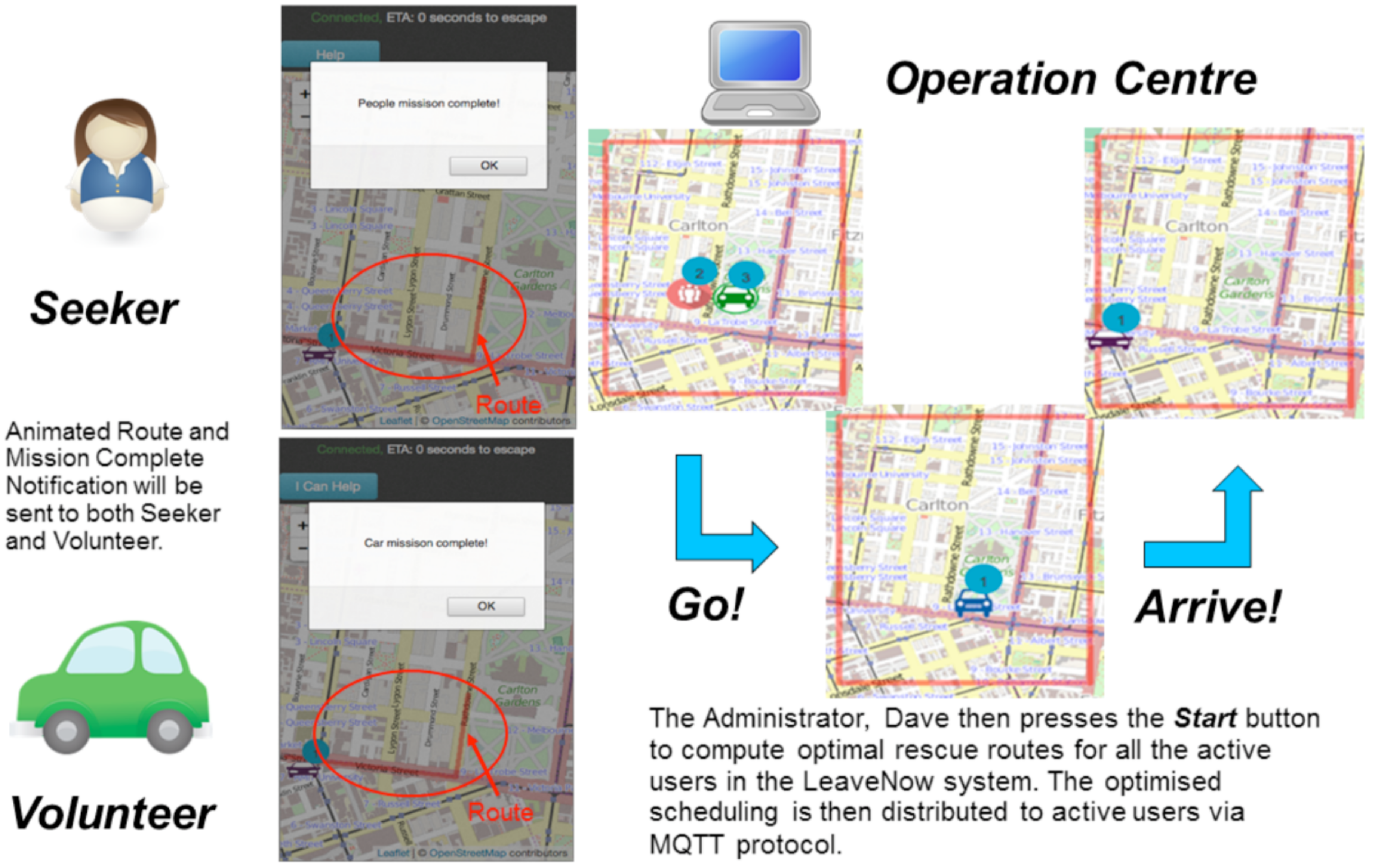}
\caption{LeaveNow user interfaces and workflow}
\label{fig:demo}
\vspace{-8pt}
\end{figure}

%\section{LeaveNow Demo}
The LeaveNow Seeker and Volunteer mobile clients gather relevant information (e.g. location, number of available seats), which is shared over the social network and is transported via the MQTT protocol~\footnote{http://mqtt.org/} supported by the MessageSight virtual appliance~\footnote{www.ibm.com/software/products/en/iot-messagesight}. The information is encapsulated and transmitted to the backend service via REST API~\footnote{REpresentational State Transfer Application Programming Interface}. The backend services optimize all requests and available resources in the rescue-and-routing scheduling. The results are then sent back to all clients in real-time, as well as made available to emergency services to keep an updated status of the rescue situation. Figure~\ref{fig:demo} shows the LeaveNow user interfaces and workflow.

\section{Evacuation Time Analysis}
\begin{figure}[!hp]
\centering
\includegraphics[width=7.5cm]{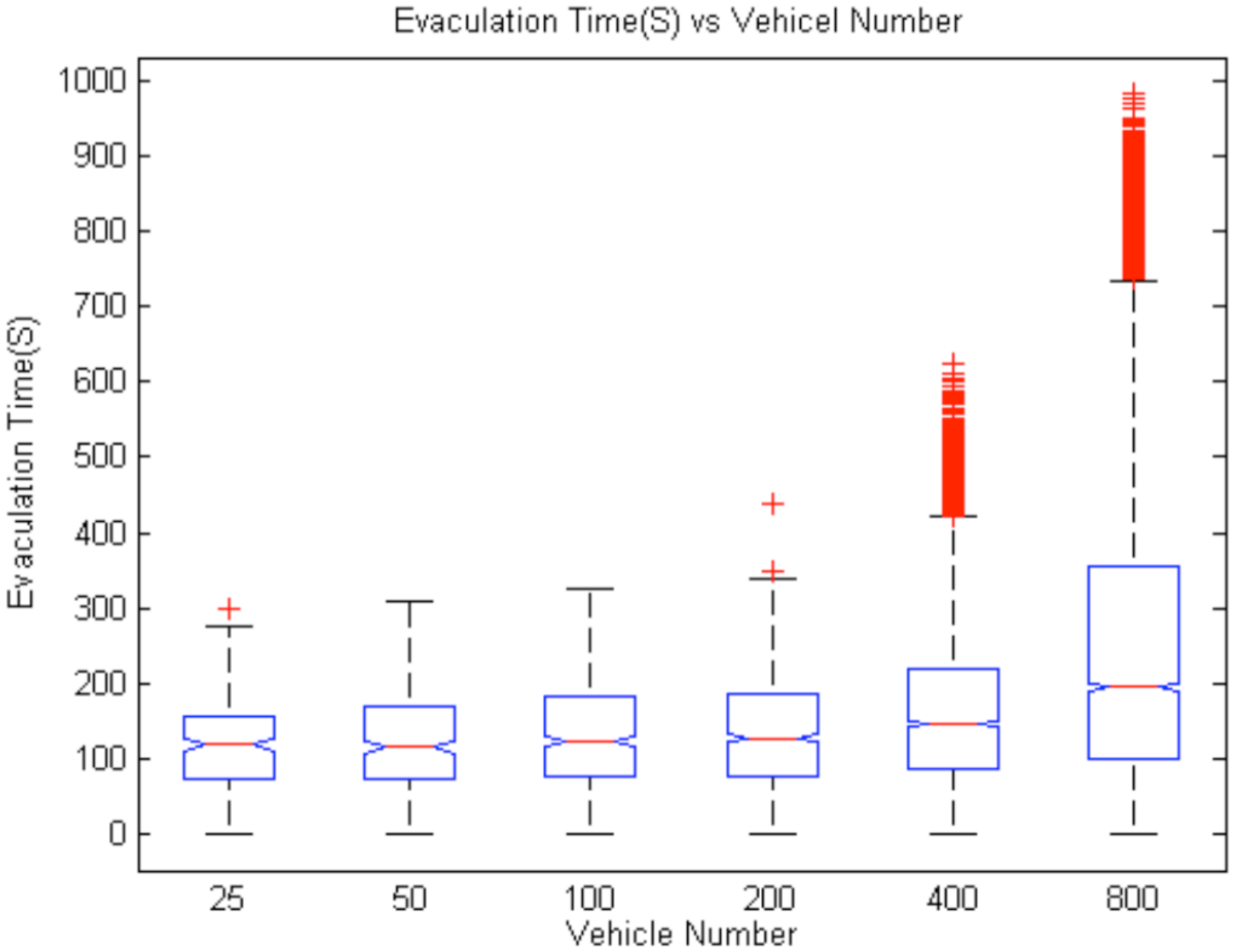}
\caption{Evacuation time analysis.}
\label{fig:results}
\vspace{-8pt}
\end{figure}

LeaveNow maximizes the number of people per vehicle and recommends users, via the social network, to use volunteer rescue vehicles instead of driving their own. Figure~\ref{fig:results} shows the evacuation time with different vehicle numbers for a danger zone size of about $30 km^2$. The average evacuation time increases linearly with the vehicle number, however, when the vehicle number reaches 400 the variation of evacuation time increases exponentially. Therefore, by reducing the number of on-road vehicles, LeaveNow mitigates congestion and leads to faster evacuation. 

\section{Conclusions \& Future Work}
LeaveNow integrates a real-time location-based social network into the coordination and optimization of evacuations during a disaster.  Via the efficient car-sharing and optimized routing services, LeaveNow reduces the load on public emergency services and coordinates social effort to save lives.
%By integrating real-time location-based social network into evacuation during disasters, we can improve efficient car-sharing coordination by optimizing reactions, as well as coordination of social effort and public emergency services to save lives. 
We plan to extend the dynamics of road networks during disasters to traffic simulations, e.g, a fallen tree branch or power line blocking the road. Thus the next version of LeaveNow will achieve a higher level of realism in computing evacuation routes and assigning seekers to volunteers. We also plan to improve the car-sharing algorithm. For example, if a seeker is not on a volunteer's route but is nearby, the volunteer can still pick up the seeker, provided it is safe to do so. This strategy is expected to result in further reduction of the number of seekers who would not be assigned for pick-up if they were not on any route of volunteers. 
%We plan to extend the way in which a seeker and a volunteer get assigned in LeaveNow. Other than the operation center makes decisions of pick-up assignment, seekers and volunteers can make assignment decisions by themselves.
%The system can be applied in other scenarios, such as car-pooling in daily life. 
%We can extend LeaveNow to a mobile app that enables people much easier and more convenient to find others to share a car to their destinations. Such a mobile app can encourage more people to take part in car-pooling.


  % sigproc.bib is the name of the Bibliography in this case
% ACM needs 'a single self-contained file'!
%

\end{document}